\documentclass[12pt]{article}
\usepackage{graphicx}
\usepackage{url}

\newcommand\pubnumber{NuPhys2015-Soleti}
\newcommand\pubdate{\today}

\def\oxford{Department of Physics\\
University of Oxford, OX1 3RH Oxford, UNITED KINGDOM}

\def\Title#1{\begin{center} {\Large #1 } \end{center}}
\def\Author#1{\begin{center}{ \sc #1} \end{center}}
\def\Address#1{\begin{center}{ \it #1} \end{center}}

\newcommand\pubblock{\rightline{\begin{tabular}{l} \pubnumber\\
         \pubdate  \end{tabular}}}
\newenvironment{Abstract}{\begin{quotation}  }{\end{quotation}}
\newenvironment{Presented}{\begin{quotation} \begin{center} 
             PRESENTED AT\end{center}\bigskip 
      \begin{center}\begin{large}}{\end{large}\end{center} \end{quotation}}
\def\Acknowledgements{\bigskip  \bigskip \begin{center} \begin{large}
             \bf ACKNOWLEDGEMENTS \end{large}\end{center}}

\def\beq{\begin{equation}}
\def\eeq#1{\label{#1}\end{equation}}
\def\eeqn{\end{equation}}

\def\beqa{\begin{eqnarray}}
\def\eeqa#1{\label{#1}\end{eqnarray}}
\def\eeqan{\end{eqnarray}}

\let\bar=\overbar

\def\Dslash{\not{\hbox{\kern-4pt $D$}}}
\def\dslash{\not{\hbox{\kern-2pt $\del$}}}

\def\msb{{\bar{\ssstyle M \kern -1pt S}}}

\begin{document}
\begin{titlepage}
\pubblock

\vfill
\Title{The Muon Counter System for the MicroBooNE experiment}
\vfill
\Author{Stefano Roberto Soleti}
\Address{\oxford}
\vfill
\begin{Abstract}
The MicroBooNE experiment is a liquid argon TPC experiment designed for short-baseline neutrino physics, currently running at Fermilab. Due to its location near the surface, cosmic muons can be a source of backgrounds to many analyses and having a good understanding of the cosmic rays will be very valuable for the experiment. These proceedings describe the physics motivation, setup, and performance of a small external muon counter system, which will provide improved calibration for the liquid argon TPC and better understanding of the cosmogenic background.
\end{Abstract}
\vfill
\begin{Presented}
NuPhys2015, Prospects in Neutrino Physics\\
Barbican Centre, London, UK,  December 16--18, 2015

\end{Presented}
\vfill
\end{titlepage}
\def\thefootnote{\fnsymbol{footnote}}
\setcounter{footnote}{0}

\section{The MicroBooNE experiment}
MicroBooNE is a short-baseline neutrino detector, consisting of a 170 ton (86 active) Liquid Argon Time Projection Chamber (LArTPC) located in the Fermilab Booster Neutrino Beamline (BNB). Its main physics goal is to address the nature (electron or photon) of the low-energy excess observed by MiniBooNE \cite{miniboone}. MicroBooNE will provide very good separation between electrons and photons, as demonstrated by the ArgoNeuT detector \cite{argoneut}.

The LArTPC is located on the axis of the 8 GeV Booster Neutrino Beamline at Fermilab, as was its predecessor MiniBooNE. The corresponding L/E ratio is $\sim1$~m/MeV. It also sits off-axis from the 120 GeV Main Injector (NuMI) beam. MicroBooNE finished commissioning in summer 2015 and has been collecting neutrino data since October 2015.

\begin{figure}[htb]
\centering
\includegraphics[height=2.5in]{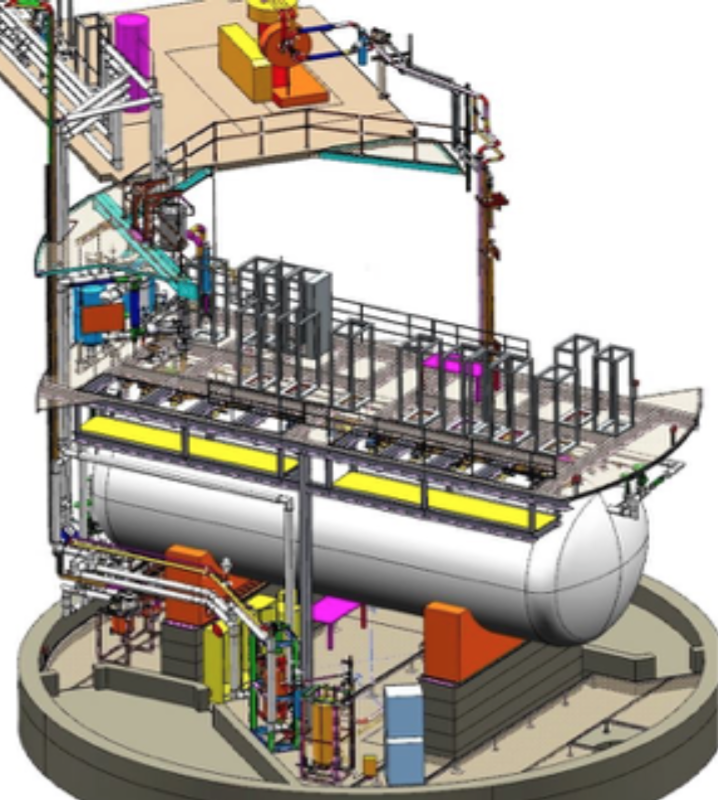}\hspace{2em}
\includegraphics[height=2.5in]{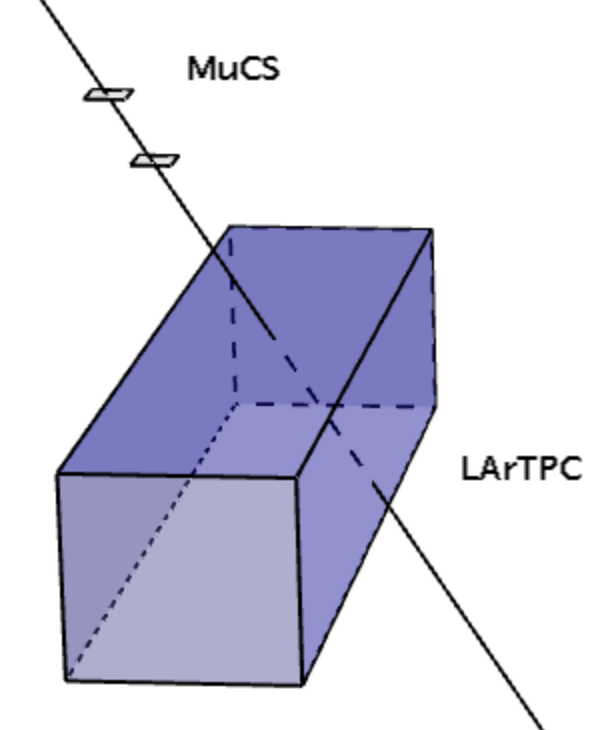}
\caption{Left: 3D drawing of the material surrounding the detector. Right: 3D drawing of the position of the MuCS with respect to MicroBooNE. The neutrino BNB beam goes through the front face of the detector.}
\label{fig:3d}
\end{figure}

\section{The Muon Counter System}

The cryostat containing the TPC is located just below the surface as shown in Fig. \ref{fig:3d} (left). The cosmic rays hitting the detector are one of the main experimental backgrounds and a good understanding of them is essential for the experiment. A small external muon counter system (MuCS) is used to obtain a clean dataset of cosmic rays, useful for several physics, calibration, and reconstruction studies and reconstruction efficiencies. The MuCS consists of two identical muon detectors, placed above the TPC, outside of the cryostat (Fig. \ref{fig:3d} right).

\begin{figure}[htb]
\centering
\includegraphics[height=3.6in]{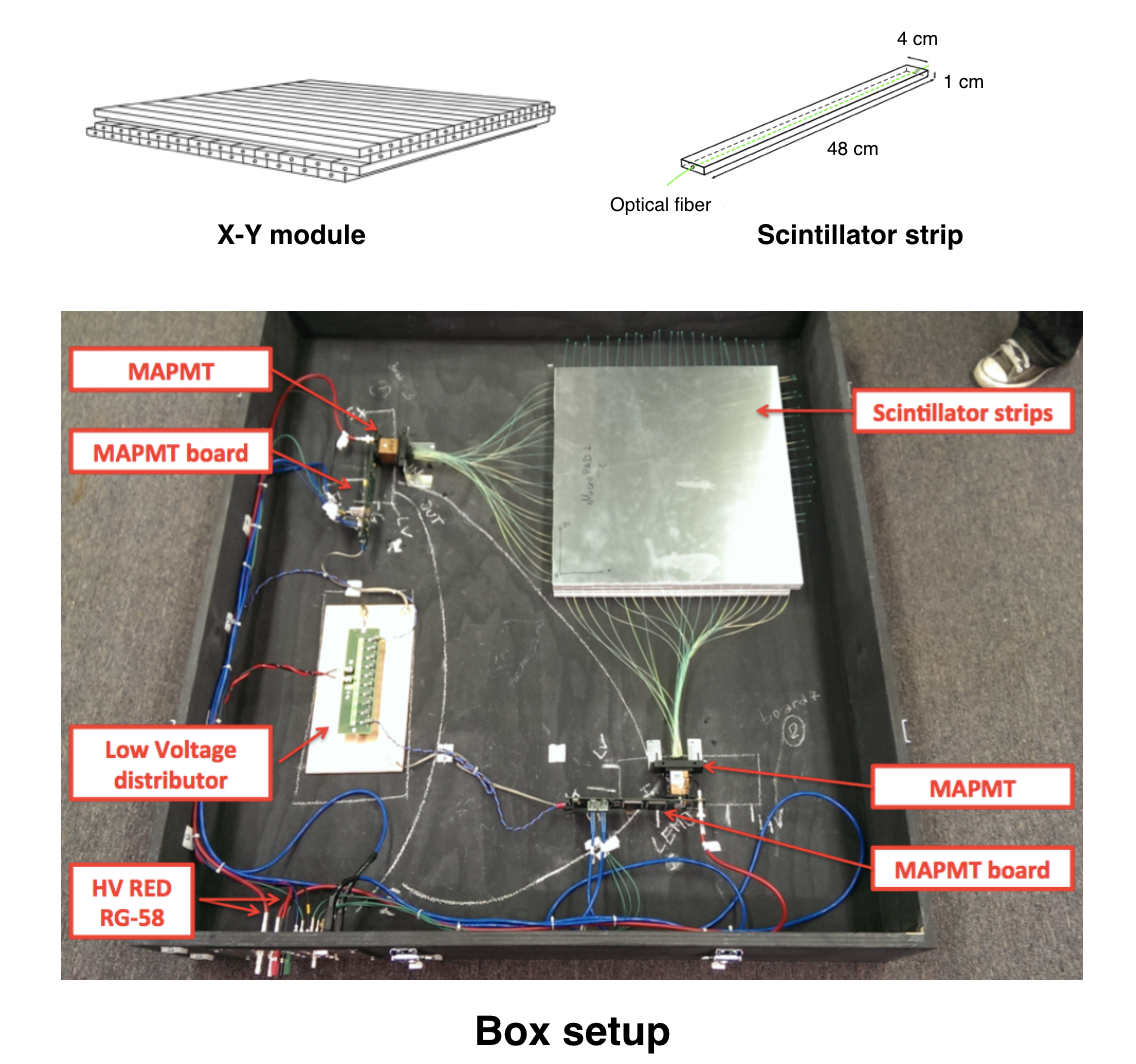}\hspace{2em}

\caption{Top: 3D drawing of a bilayer and dimensions of a scintillator strip. Bottom: Picture of the setup of a MuCS detector, with the two bilayers in the top-right corner.}
\label{fig:boxsetup}
\end{figure}

Each detector is made of two bilayers of scintillator strips 48x4x1 cm$^{3}$, placed one above the other. Each bilayer is oriented parallel to the beam direction and perpendicular to each other (Fig. \ref{fig:boxsetup}, top). This disposition enables the measurement of two-dimensional ($x,y$) spatial information for through-going cosmic rays. The staggered positioning of the bilayers gives a spatial resolution of 2 cm on both axis. Light is collected by optical fibers and readout by two separate multi-anode photomultipliers (MAPMTs) Hamamatsu H8804 \cite{hamamatsu}, as shown in Fig. \ref{fig:boxsetup} (bottom).

\subsection{Cosmic ray extrapolation}
In MicroBooNE, timing information is provided by a 32 photomultiplier tube (PMT) array, measuring the prompt scintillation light produced during the excitation or ionization of argon atoms. The flash times are then matched to the reconstructed tracks offline. 

It is then possible to associate a track in the TPC to the hits in the MuCS in two ways:
\begin{itemize}
\item comparing the hit time in the MuCS with the flash time matched to the track;
\item extrapolating the cosmic ray path from the starting point and direction of the track in the TPC and check if it is close to the hits in the counters.
\end{itemize}

The location of the boxes allows to tag cosmic rays covering  $\sim$43\% of the TPC volume. The fraction of cosmic rays passing through the Muon Counter System without hitting the TPC is, in the current configuration, 0.3\%. 

\subsection{Cosmic ray spatial distribution}
Data collected by the MuCS can be used to measure the angular and spatial distributions of the cosmic rays hitting both boxes. For MuCS data, the MicroBooNE DAQ is set to be triggered by a muon that generates a 4-bilayer coincidence.

Knowing the strips hit by the cosmic ray it is possible to extrapolate two sets of ($x,y$) coordinates, one for each box, and measure the polar and azimuthal angles of the cosmic ray. Comparing data to Monte Carlo, the relative positioning of the boxes can be checked and compared with alignment measurements.

\begin{figure}[htb]
\centering
\includegraphics[height=1.9in]{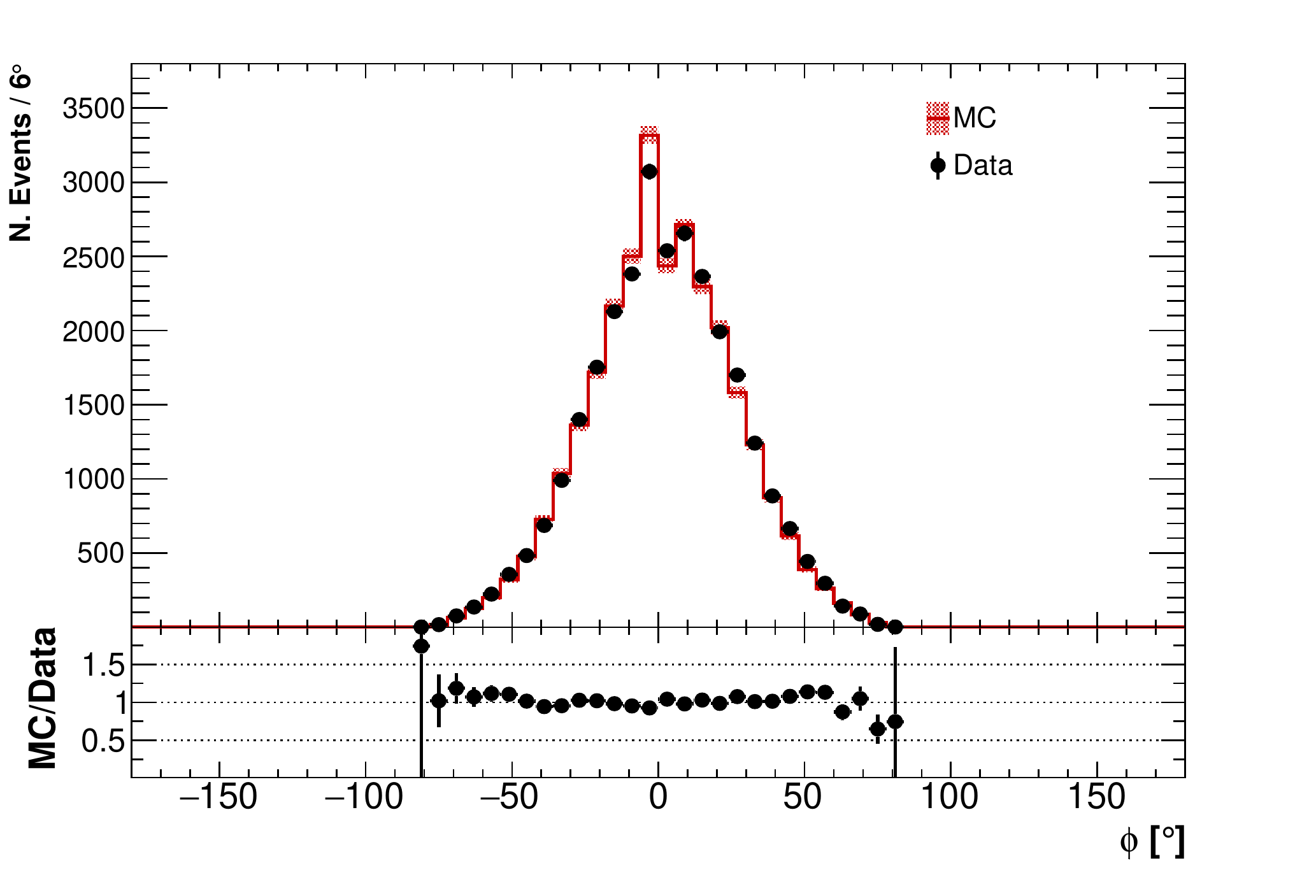}
\includegraphics[height=1.9in]{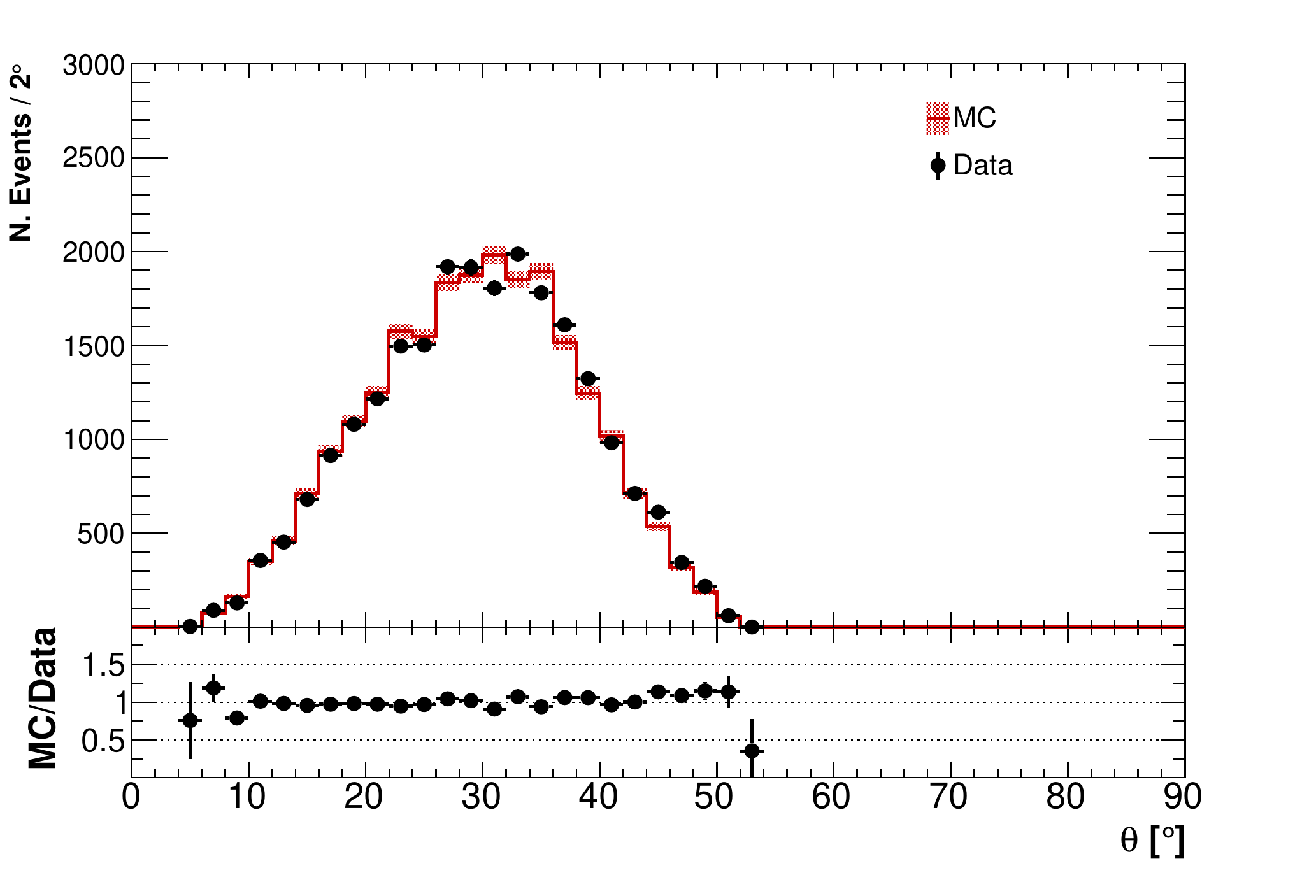}
\includegraphics[height=1.9in]{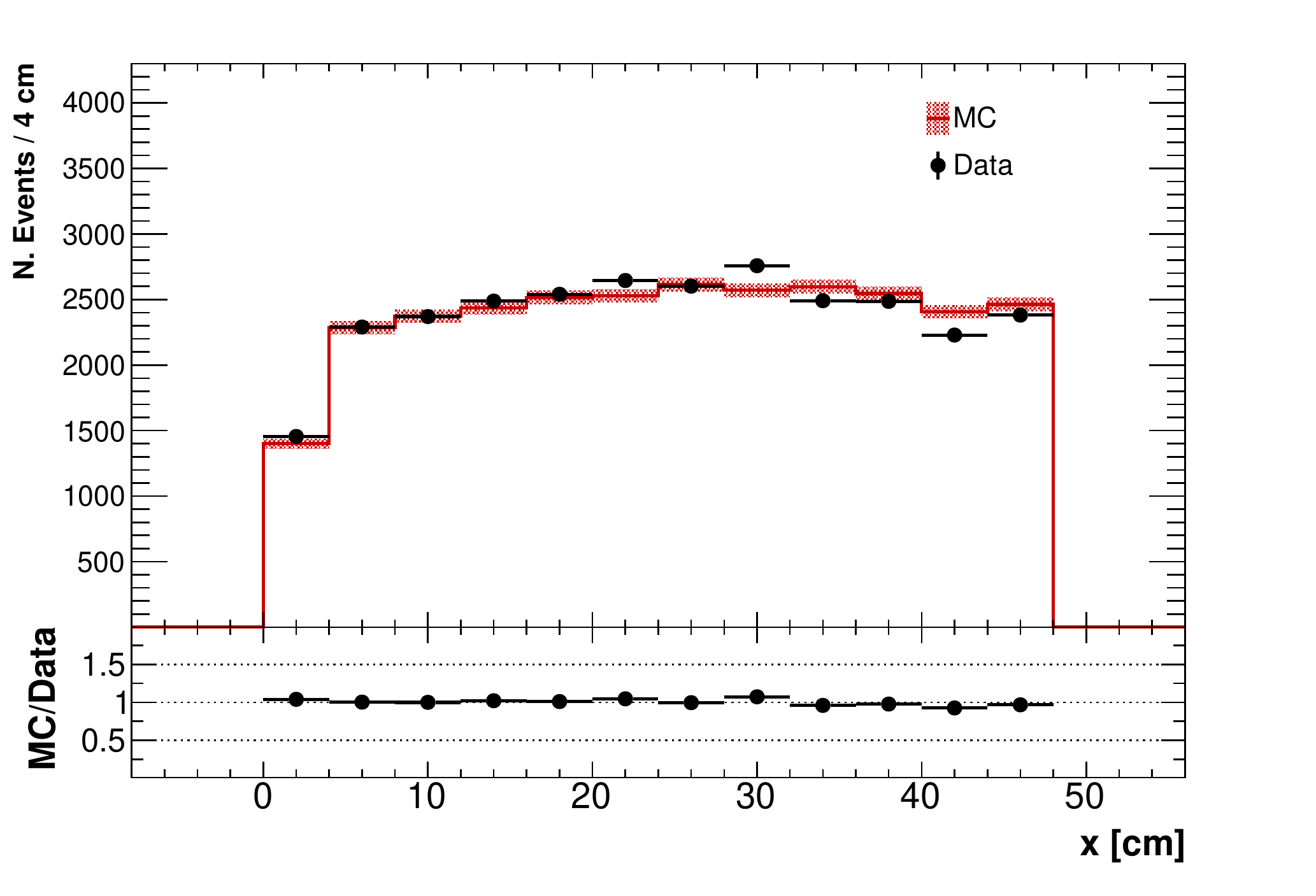}
\includegraphics[height=1.9in]{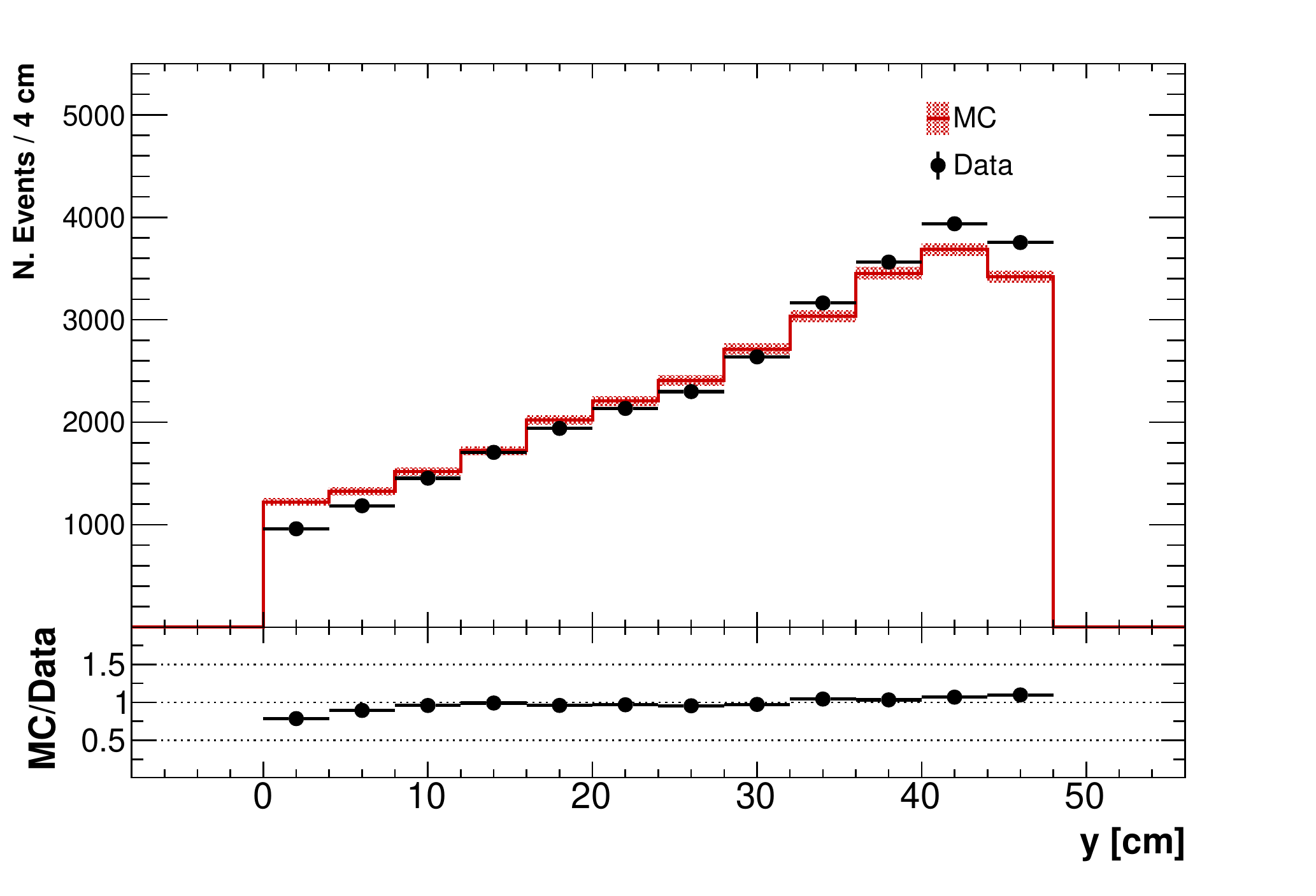}
\caption{Data and Monte Carlo distributions for angular distribution (top) and starting points coordinates (bottom).}
\label{fig:data}
\end{figure}

The distribution of the starting $y$ coordinate (in this system it corresponds to the direction perpendicular to the beam) shows an increasing slope because the two boxes are shifted on this axis (Fig. \ref{fig:data}, bottom right). The distribution of the starting $x$ coordinate is constant, except for the lower first bin, due to the slight shift of the two boxes in this direction (Fig. \ref{fig:data}, bottom left).
The azimuthal angle $\phi$, measured starting from the $y$ axis, is peaked around 0$^{\circ}$ as expected, because of the alignment of the boxes in the $y$ direction (Fig. \ref{fig:data}, top left). The polar angle $\theta$, measured from the zenith, is peaked around 30$^{\circ}$, because of the horizontal shift of the two detectors (Fig. \ref{fig:data}, top right). 
The agreement between data and Monte Carlo is good for all the four distributions.

\subsection{Analysis topics}
Once a clean dataset of cosmic muons passing through the MuCS has been obtained it will be possible to perform several physics analysis and detector studies:
\begin{itemize}
\item \emph{trigger efficiency}. Comparing the flash time with the MuCS time could allow to cross-check the trigger efficiency;
\item \emph{data reconstruction efficiency}. A clean dataset of cosmic rays can be used to compare the reconstruction efficiencies of different algorithms;
\item \emph{detector performance}. Comparing the measured collected charged in the TPC and the measured number of photoelectrons in the optical system with the expected ones could allow to measure detector performances;
\item \emph{calibration studies}. Minimum ionizing particles can be used to cross-check liquid argon purity and PMT gain and quantum efficiency measurements.
\end{itemize}

The MuCS has been commissioned and has started taking data. Analysis is already underway with the acquired samples to perform these tasks.


\Acknowledgements
The author thanks the organizers of NuPhys 2015 conference for accepting my contribution to the poster session. Special thanks goes to R. Guenette, M. Bass and G. Zeller for their help in improving this work.

\end{document}